\documentclass[twocolumn,prl]{revtex4}
\usepackage{graphicx}

\def\r{{\bf r}}
\def\k{{\bf k}}

\begin{document}

\title{Polymers in a vacuum}

\author{J. M. Deutsch}
\affiliation{ Department of Physics, University of California, Santa Cruz, CA 95064.}

\date{\today}

\begin{abstract}
In a variety of situations, isolated polymer molecules are found
in a vacuum and here we examine their properties. Angular momentum
conservation is shown to significantly alter the average size of a chain and its
conservation is only broken slowly by thermal radiation. The time autocorrelation for monomer
position oscillates with a characteristic time proportional to chain
length. The oscillations and damping are analyzed in detail.
Short range repulsive interactions suppress oscillations 
and speed up relaxation
but stretched chains still show damped oscillatory time correlations.
\end{abstract}


\maketitle

The properties of polymer chains have been investigated extensively over the
past fifty years~\cite{degennes} but
the vast majority of these studies have been concerned with situations where
they are in a solution or a melt.  However there are some situations
where polymer molecules are essentially in a vacuum. Desorption and ionization
of polymers, often by lasers is carried out during mass spectrometry, in order to 
characterize desorbed proteins. This
has many important applications including the understanding of
cancer~\cite{CaldwellCaprioli}. Polymer molecules of many different kinds 
have been detected in interstellar media~\cite{NewCarbonChains} and
although they have predominantly had less than 10 units, the
detection of new species is an active area of research. It might also
be possible to employ optical tweezers on biomolecules such as DNA
and manipulate them in a vacuum, in a manner similar to what is now done 
routinely in aqueous solution~\cite{block}.

To aid in the possible experimental observation of such systems, some
basic properties of isolated polymers in a vacuum are considered here.
The first question that we ask is how their statistics are modified from
those in solution. Solvents will compete with intra-chain attractions
so that above the $\theta$ temperature~\cite{degennes}, a polymer chain will be
swollen. Without the solvent present, this would imply that a chain at
the same temperature would be collapsed. But at high enough temperatures,
entropy will dominate over energy, and a polymer, just like a liquid,
will then want to expand into a gas, or self avoiding phases. Because carbon-carbon bonds are very strong, it
might then be possible to find some species where a polymer will become
swollen in isolation for long enough periods of time to
be observable. Even if it turns out that this is not possible,
polymers through desorption often carry a charge, and this additional
coulomb repulsion is quite substantial, at $500 K$ it is $\approx 33 k_B T$
for two electrons $1 ~nm$ apart. This will serve to stretch a chain. Such a situation is a
possibility in the desorption and ionization of proteins done
in mass spectrometry experiments\cite{CaldwellCaprioli}.

It might then appear that the statistics of this system are identical to
that of a chain in a solvent, with some modification of interaction
parameters. However one important difference is the conservation of
angular momentum that we might expect to see in this case as
opposed to a polymer in a solvent. In statistical mechanics, this conservation law 
is ordinarily ignored and is
not expected to make a difference to system properties when the number
degrees of freedom are large.  However we will see that for a polymer
in isolation it has a significant effect on its size, even when the total
angular momentum is zero.
The effects of angular momentum conservation have been recently 
studied in self-gravitating systems~\cite{laliena} 
where it leads to different phases for some models for finite angular
momentum.

The  starting point for this situation is the formula for the classical entropy in
the microcanonical ensemble with
conservation of linear and angular momentum of a system with total 
potential energy $U$~\cite{laliena}.
However we can safely transform this into the canonical ensemble
at temperature $T$
using the usual argument that the fluctuations in the energy at 
constant temperature are small for a large number of degrees
of freedom. However we are still keeping the $\delta$ function  constraint
on the total angular momentum $\bf L$.
Using the Fourier representation for a $\delta$ function and
integrating over the momenta, the partition function becomes

\begin{equation}
Z(T,{\bf L}) \propto \int e^{i{\k}\cdot {\bf L}} e^{-\frac{T}{2}(\k \cdot {\bf I} \cdot \k) -\beta U} \prod_{i=1}^N d^3{\bf r}_i d^3 {\bf k} 
\end{equation}
where $N$ is the number of monomers, $\bf I$ the moment of inertia tensor of a polymer conformation, 
and the center of mass is fixed to zero.
Adding a term $\epsilon \sum_i r_i^2$ to $\beta U$ allows one
to differentiate $\ln Z$ with respect to $\epsilon$ to obtain the
r.m.s. size of a chain defined as
$
\langle R^2\rangle \equiv \frac{1}{N}\langle \sum_{i=1}^N r^2_i\rangle
$

\begin{figure}[t]
\begin{center}
\includegraphics[width=0.9\hsize]{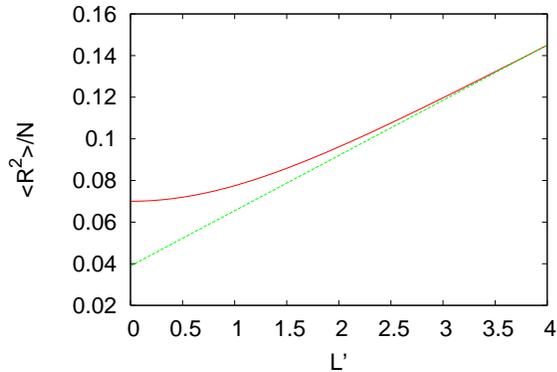}
\caption{ (Color Online) 
The variation of the size of a chain versus
its angular momentum for a ideal ring chain.
The straight line is the asymptotic slope which
can be obtained through a simple analysis.
}
\label{fig:rsq_vs_L}
\end{center}
\end{figure}

For an ``ideal" or  ``phantom" chain~\cite{footnote:ideal}  with ring topology, the 
calculation can be done exactly~\cite{DeutschToBePublished}
giving the results shown in Fig. \ref{fig:rsq_vs_L}. The rescaled
angular momentum $L' \equiv L\sqrt{12}/(N\sqrt{T m l})$ where $m$ is the mass
and $l$ is the step length, which both can be taken equal to $1$.

At $L=0$, $\langle R^2\rangle/N = .07$. Results using a simulation
method
described below give $.071$ which are the same to within the error
bars. This is substantially below, $1/12 \approx .083$, the value of 
the same quantity where conservation of angular momentum is not enforced.
When averaged over all angular momenta, the size of the chain must
agree with the non-conserved case, and because high $L$ chains will have 
greatly extended conformations, this must be compensated for by correspondingly
compact configurations for small $L$.

To obtain the asymptotic behavior for large $L$, using
a much simpler argument, we expect that in this limit
the dominant configuration of the chain will be a highly 
stretched circle of radius $R$ rotating symmetrically
about the axis of angular momentum.  We minimize the free energy,
of a polymer taking into account both its kinetic energy
and elastic energy
yielding $R^2/N = L' /(12\pi)$ again choosing $l=m=1$. 
The straight line in fig. \ref{fig:rsq_vs_L} has
the same slope.

The case of self avoiding (swollen) chains 
($R \sim N^\nu$ with $\nu \approx 3/5$), is qualitatively
similar. and the asymptotic scaling can be easily worked out
along similar lines as above, giving $R \propto L^{4/11}N^{1/11}$.

The total angular momentum however, is {\em not} conserved. Interaction
with thermal photons will cause the angular momentum to equilibrate
on a time scale that we will now estimate.
First we consider the flux of electromagnetic energy emitted
by a single polymer. The emission of thermal radiation per
unit area of a black body is given by the Stephan-Boltzmann
law $S = \sigma T^4$, where $\sigma$ is the Stephan-Boltzmann
constant. However this greatly overestimates the radiation
because of the weak efficiency of small objects in emitting
light of a far greater wavelength. Calculations for metal nanoparticles
(which should be better emitters than dielectrics)
give a suppression factor of less than $10^{-3}$ when the
nanoparticles are $50 A$ in radius \cite{radiation_small}.
This gives a ratio of  $k_B T$ to emitted power of $6\times 10^{-6} s$.
The relaxation time as we will see below scales roughly as $N^2$.
The microscopic hopping time typical for such a system is
$10^{-12}s$, meaning that the relaxation time for $N=100$ is roughly $10^{-8} s$.
Thus thermal photon equilibration is more than two orders of magnitude slower 
than the time-scale for relaxation of a chain.
Therefore one expects to see transitions in the time averaged radius
of gyration of a chain as photons are emitted and absorbed by the
polymer. This might be observable in the signature of noise seen in
light scattering.

We now turn to a study of the dynamics of these polymers and we will see
that in this respect the situation is very different from that of a polymer
in a solvent. In a solvent, models for dynamics for the most part consider
monomers connected together by linear springs, thermal noise, solvent
drag, and hydrodynamic interactions.  However in a vacuum,
the last three terms are not present, which leaves us with
Newtonian dynamics for linear springs, but
this will never thermalize, so nonlinear forces must be considered.

One might first guess that sufficient nonlinearity would introduce strong
enough dissipation of individual modes so that the behavior would be
similar to that of the Rouse model~\cite{Rouse}, which describes a ``free draining"
chain. That is one where 
individual monomers experience a drag proportional to their velocity.
However this violates Galilean invariance, as we shall discuss in
more detail below.
Moreover one dimensional systems are notorious for not being able to
equilibrate energy well, and there are many well studied instances
where this problem is known to occur. 
Perhaps the best known example
of this is the Fermi Pasta Ulam chain~\cite{FPU}. 
For small enough energies this system shows strong
recurrences in amplitude of modes above which it equilibrates~\cite{BermanIzrailev}
Models that are even more nonlinear
such as the Sinai-Chernov ``Pen Case model"~\cite{pencase} 
do not suffer from this problem,
however they still exhibit highly non-local time correlations, with universal
power law decays~\cite{DeutschNarayan},  which is a general result for one dimensional chains that are
momentum and energy conserving~\cite{NarayanRamaswamy}

So we first consider the dynamics of a chain neglecting any self avoiding
interactions, but using an athermal highly nonlinear model for the reasons
mentioned above.
Thus we have chosen a model where monomers of equal mass are coupled together by
links of fixed length. Aside from this constraint, there is
no potential energy. The monomers can freely rotate
but there is no coupling to an outside system so that
there is no dissipation or random noise term. The
model rigorously satisfies conservation of energy, momentum
and angular momentum. An efficient method
for evolving such chains was developed so that despite the
large number of length constraints, the computation for each 
time step scales linearly with the number of monomers. The 
details will be published elsewhere~\cite{DeutschToBePublished}.
The angular
momentum, center of mass and energy were monitored to ensure
that their drifts due to numerical error remained small for
all data used.

\begin{figure}[t]
\begin{center}
\includegraphics[width=\hsize]{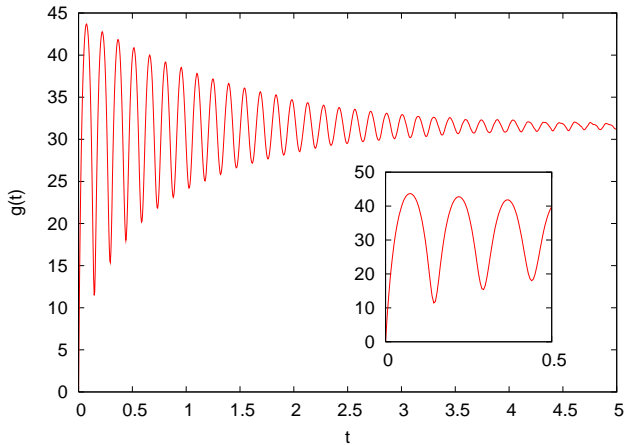}
\caption{ (Color Online) 
The autocorrelation function $g(t)$ as defined
in Eq. (\ref{eq:g(t)}) for a chain of 128 monomers. The inset shows
the beginning of the plot at a higher magnification, where the
initial linear increase and cusps are apparent.
}
\label{fig:g_vs_t}
\end{center}
\end{figure}

The monomer-monomer autocorrelation function defined as
\begin{equation}
g(t) = \langle \frac{1}{N} \sum_{i=1}^N |\r_i(t) - \r_i(0)|^2\rangle
      \label{eq:g(t)}
\end{equation}
was calculated for chains of different lengths and is displayed in
Fig. \ref{fig:g_vs_t} for $N=128$ averaged over $24,000$ runs. This is very unlike the correlation function
for a polymer in solution which shows a smooth slow increase, not
the wildly oscillatory form seen here. The period of these
oscillations scales as chain length. This is what one would expect
for a linear model with no friction, because the lowest mode of
oscillation has spring constant $\propto 1/N$, and the mass is
$\propto N$ giving a period $\propto N$. At first sight, it is
surprising that the oscillations are so weakly damped in a model
that is so highly nonlinear. To understand this better, we analyze
this problem in terms of linear modes. The relaxation time of the chain
is then given by the damping time for these oscillations. In terms
of a Fourier decomposition along the arclength of the chain,
we can model the correlation function as
$\langle r_k(0)r_k^*(t)\rangle = \langle|r_k|^2\rangle Re(\exp((i\omega_k -\lambda_k) t)$
and because $|r_k|^2 \propto 1/k^2$, 
\begin{equation}
g(t) \propto \sum_k \frac{(1-\cos(\omega_k t))}{k^2}e^{-\lambda_k t}
\label{eqn:g(t)lambda}
\end{equation}
and this can be used to fit the numerical data. In this expression the $\omega_k$'s represent
the frequencies of oscillation.  We therefore expect that for small $k$ that $\lambda_k$ will 
be small in comparison to $\omega_k$, the only other relevant frequency scale
for each term in the summation.

For small $t$, and no dissipation ($\lambda_k=0$), the
small $t$ behavior of $g(t)$ in the above expression can easily be shown to be $\propto t$. 
Yet for short times this shows that the mean square displacement follows the same law as a diffusive
process. For longer times, oscillatory behavior will be seen but will be asymmetric,
showing cusps at the minima and parabolic maximum. These same qualitative
features persist for small $\lambda_k$ and are also in quite good agreement
with the simulation data shown in Fig. \ref{fig:g_vs_t}.

It is clear from the data that $\omega_k \propto k$, just as one would expect
from a linear system of springs. The damping appears to fit best to
a form close to $\lambda_k \propto k^2$. Fitting this to different
chain lengths, $N=64$ and $128$, gives a relaxation time 
$T_{rel} \propto N^{(1.85\pm.15)}$. Note that in
the case of one dimensional
heat conduction, it has been found that even with highly
nonlinear models~\cite{DeutschNarayan}, asymptotic large $N$ behavior is difficult
to study as more than $10^4$ particles must be considered to
get a good estimate of critical exponents. Therefore it is
possible that the exponent found is off by $\sim 10\%$ of its asymptotic
value.

This value of the relaxation time scales closely to what is seen in the Rouse
model although in that case the physics is very different as there is no inertia term
and no rapidly oscillatory behavior. In the case we consider here, the
origin of this time is due to nonlinear coupling of different modes 
resulting in the slow translation to decoherence 
as time progresses. 

As mentioned above,
this problem is quite similar to that of
a one dimensional nonlinear chain of particles.
In that case the system is also
characterized by long wavelength oscillations that slowly decohere. But there,
the relaxation time $T_{rel} \propto N^{3/2}$ which is different than the
polymer case. If the polymer chain was stretched by a constant force
so that it was quasi-one dimensional, one would expect the same $N^{3/2}$ scaling
for the relaxation time. 

To understand the damped oscillatory behavior of the correlation
function, in Eq. \ref{eqn:g(t)lambda}, we
consider what kind of coarse grained 
linear stochastic equation would best approximate the evolution of ${\bf r}(s,t)$,
the position of the chain at arclength $s$ and time $t$.
Because the system has Galilean invariance, there can be no
$\dot {\bf r}$ term for the damping, as center of mass velocity
is conserved. By symmetry, the lowest order damping term must be
$C \partial^3 {\bf r}/\partial t\partial^2 s$ where C is a constant.
Adding in inertia, random forcing $\bf\xi(s,t)$ and chain
connectivity gives
\begin{equation}
\frac{\partial^2 {\bf r}}{\partial t^2} =
\Big(1 + C \frac{\partial}{\partial t} \Big) \frac{\partial^2 {\bf r}}{\partial s^2}  +
{\bf \xi(s,t)}
\label{eq:rouselike}
\end{equation}
In relation to the above analysis, this gives
a damping $\lambda_k \propto k^2$ for small k.
Such a model matches fairly well the numerical data~\cite{footnote:RelaxTime}.

\begin{figure}[t]
\begin{center}
\includegraphics[width=\hsize]{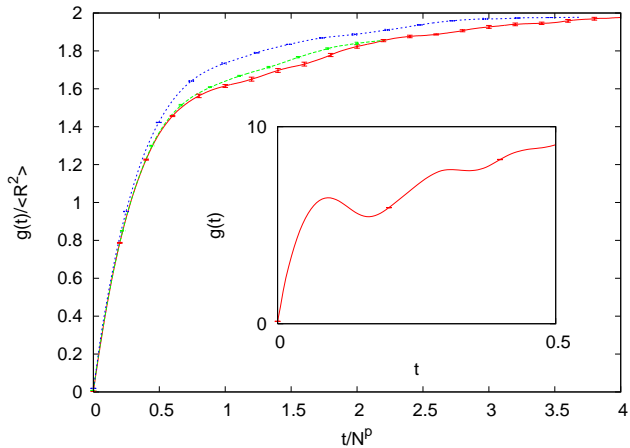}
\caption{ (Color Online) 
The scaled autocorrelation function $g(t)/\langle R^2\rangle$ versus
$t/N^p$, for three chain lengths, $N=32$, $64$, and $128$ with
short-range repulsive interactions. The inset shows $g(t)$ for an $N=32$ chain
with equal charges at both ends that cause it to stretch.
}
\label{fig:g_excl}
\end{center}
\end{figure}

Such a proposal for internal damping for polymer chains in connection with
Cerf friction~\cite{degennes,Cerf} has been made before~\cite{MacInnes} using a nonrigorous derivation 
that gives rise to the same third order derivative term as in eqn. \ref{eq:rouselike}.
Adding such a term to the Rouse equation provides an explanation
of experiments~\cite{CerfExp} on extensional relaxation of polymers in solvent.
Solvents with different viscosities were considered to extrapolate to 
the limit of zero viscosity, and the results can be interpreted using
such a term.

As one might expect, the inclusion of repulsive interactions between
monomers suppresses the oscillations that are seen in the ideal chain.
For the sake of efficiency,
soft-core potentials were added between between monomers having
a potential of the form $\beta V(r) = 2(1-(r/l)^2)^5$. 
Statistics of
such chains with total angular momentum of zero were measured 
and the size scaling exponent gave $\nu = .596 \pm .01$ in 
good agreement with the well known three dimensional value.

In Fig. \ref{fig:g_excl}
the autocorrelation function for these chains, see Eq. \ref{eq:g(t)}, 
is plotted for chain lengths $N=32$, $64$, and $128$, on scaled axes so
that they coincide for short times. The vertical axis is $g(t)/\langle R^2\rangle$
and the horizontal one is $t/N^p$, with $p$ chosen to fit short times
best. With $p = 1.15$, the plots coincide well over half of the vertical
range, from $0$ to $1$. However the long time behavior for $N=32$
is noticeably above the longer length chains. However $N=64$ is only
slightly above $N=128$, and given the correlated error bars, this is
barely statistically significant. This is strong evidence that for
large $N$ the correlation function approaches the scaling form $g(t)
= N^{2\nu}f(t/N^p)$, and therefore the relaxation time for this chain is
$\propto N^p$, with $p = 1.15\pm .05$. Note that this is much smaller than
that of the ideal chain discussed above, presumably because long range
interactions along the chain backbone allow much faster equilibration
of energy and momentum~\cite{footnote:LongerRelaxTime}.  We expect the
time it takes a chain segment to move of order its average size $R_g$ should
be $R_g$ divided by the center of mass speed for of order
half the chain, $\sim N^{-1/2}$, which gives $t_{rel} \sim N^{1.1}$.

Finally we contrast this with what happens if charges are added to both ends.
With charged protein molecules observed in mass spectrometry, a similar situation
could also occur. The inset in Fig. \ref{fig:g_excl} shows the autocorrelation
function for this case, where the end to end distance is $10.03$, about one
third of the chain's arclength. The parameters were chosen so that there is still a substantial
amount of interaction between neighboring monomers, but the chain is still quite
stretched. Here one can clearly see oscillations in $g(t)$, intermediate
in behavior between the ideal chain and interacting cases.

In conclusion, the equilibrium statistics and dynamics of polymers in a
vacuum have many interesting properties. The addition of angular momentum
conservation significantly alters chain statistics. The subtle power-law
time correlations found in momentum conserving one dimensional systems
can lead to dynamics that are oscillatory and show unusual scaling
properties. It is hoped that this work will provide impetus for further
experimental observation of these fascinating systems.

The author thanks Larry Sorensen for very useful discussions.


\begin{thebibliography}{}
\bibitem{footnote:RelaxTime}
The fact that the terminal relaxation time found scales as $N^{1.85}$
and not $N^2$, means that $C$ will have a very weak $\approx N^{0.15}$
dependence on $N$ which if actually correct would imply that there is
some nonlocality in the equation of motion. However it is
likely that this deviation is due to the finite length
of the chains considered, as experience in the one-dimensional
heat conduction problem has shown.
\bibitem{footnote:LongerRelaxTime}
One can still not
rule out the possibility of a longer relaxation time with a signature too small
to be measurable.
\end{thebibliography}
\end{document}